\documentclass[showpacs,preprintnumbers,prb,aps,preprint]{revtex4-2} 

\usepackage{float} 
\usepackage{graphicx}
\usepackage{dcolumn}
\usepackage{bm}
\usepackage{textcomp}
\usepackage{ulem}
\usepackage{multirow}
\usepackage[latin1]{inputenc}
\usepackage[english]{babel}
\usepackage{hyperref}
\usepackage{enumerate}
\usepackage{amstext} 
\usepackage{amsmath}
\usepackage{amssymb}
\usepackage{soul} 
\graphicspath{{Fig/}} 
\usepackage{color}

\usepackage{stmaryrd}

\begin{document}
\title{Long-range order in arrays of composite and monolithic magneto-toroidal moments}
\author{Jannis Lehmann$^1$}
\email{jannis.lehmann@mat.ethz.ch}
\author{Na\"{e}mi Leo$^{2,3,4,5}$}
\author{Laura J. Heyderman$^{2,3}$}
\author{Manfred Fiebig$^1$}
\affiliation{$^1$Laboratory for Multifunctional Ferroic Materials, Department of Materials, ETH Zurich, 8093 Zurich, Switzerland}
\affiliation{$^2$Laboratory for Mesoscopic Systems, Department of Materials, ETH Zurich, 8093 Zurich, Switzerland}
\affiliation{$^3$Laboratory for Multiscale Materials Experiments, Paul Scherrer Institute, 5232 Villigen PSI, Switzerland}
\affiliation{$^4$Nanomagnetism Group, CIC nanoGUNE BRTA, 20018 Donostia -- San Sebasti\'{a}n, Spain}
\affiliation{$^5$Instituto de Nanociencia y Materiales de Aragón (INMA), CSIC-Universidad de Zaragoza, 50009 Zaragoza, Spain}

\date{\today}

\begin{abstract}

\noindent
Magneto-toroidal order, also called ferrotoroidicity, is the most recently established type of ferroic state.
It is based on a spontaneous and uniform alignment of unit-cell-sized magnetic whirls, called magneto-toroidal moments, associated with a macroscopic toroidization.
Because of its intrinsic magnetoelectric coupling, this new ferroic state could be useful in the development of spintronic devices.
We exploit two-dimensional periodic arrays of magnetostatically coupled nanomagnets as model systems for the investigation of long-range magneto-toroidal order.
We present two pathways promoting this order, namely (i), structures comprising a ring of uniformly magnetized sub-micrometer-sized bar magnets and (ii), structures in which each magnetic building block itself hosts a magnetic vortex.
For both cases calculations of the magnetic-dipole interaction and micromagnetic simulations reveal the conditions for the formation of spontaneous magneto-toroidal order.
We confirm this order and the formation of magneto-toroidal domains in our arrays by magnetic force microscopy.
We identify the presence of two types of domain-wall states emerging from the competition of two intrinsic microscopic couplings.
Our work not only identifies the microscopic conditions promoting spontaneous magneto-toroidal order but also highlights the possibility tailor mesoscale magnetic arrays towards elusive types of ferroic order.
\end{abstract}

\maketitle

\section{Introduction}

\begin{figure}[H]
	\centering
	\includegraphics[width = .5\textwidth]{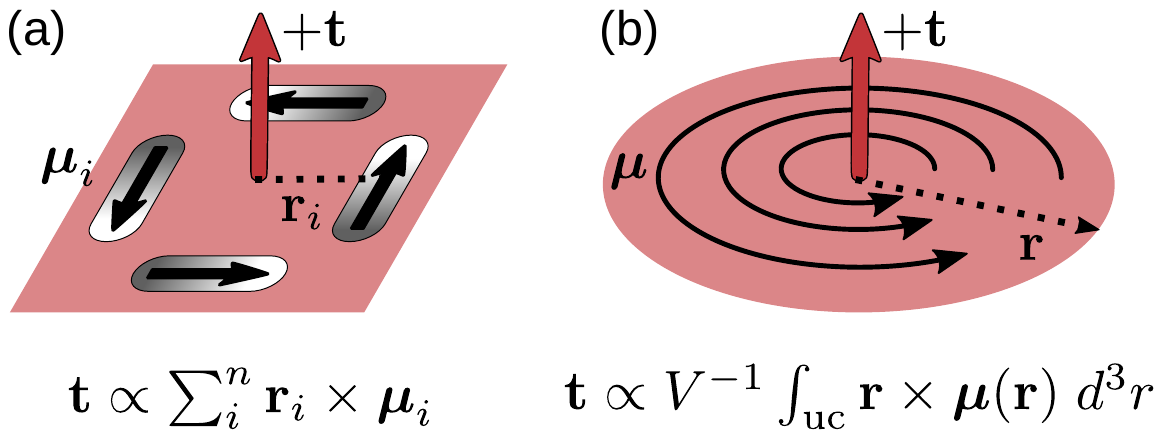}
	\caption{\label{Tisol}\textbf{Two-dimensional microscopic representations of magneto-toroidal moments.}
	(a) Composite magneto-toroidal moment $\mathbf{t}$ originating either from discrete quantum-mechanical spins or from classical magnetic moments $\boldsymbol{\mu}_i$ (black arrows and gradients) displaced by distance vectors $\mathbf{r}_i$ from the origin.
	(b) Monolithic magneto-toroidal moment originating either from localized orbital currents of atoms or ions or from a continuous vortex-like magnetization configuration within a single magnetically ordered entity.}
\end{figure}

\noindent
The search for self-organized ordered or strongly-correlated states of matter is a fascinating subject of physics and materials science.
Ferroic materials, which are related to a spontaneous and reorientable magnetic, electric or structural order~\cite{Aizu70, Wadhawan00, Tagantsev10}, are of particular interest as they provide the basis for a plethora of technological applications.
Ferroic materials are defined by the existence of a spontaneous point-group-symmetry-breaking phase transition with the formation of domains as regions described by different, yet uniform orientations of the so-called order parameter as a macroscopic observable classifying the phase transition.
This order parameter, which may be the magnetization in the case of ferromagnetism, has to be orientable by a conjugate field, which, for the magnetization, is a magnetic field.
Importantly, beyond this purely macroscopic definition, ferroic materials require microscopic interactions that support and stabilize the associated spin, charge or distortive order.
The identification and understanding of new types of ferroic states complementing the established ones (ferromagnetism, ferroelectricity and ferroelasticity~\cite{Wadhawan00}) is a task of great current interest~\cite{Spaldin13, Jin20}.
In this respect, ferrotoroidicity has recently been proposed as a ferroic state defined by the spontaneous uniform long-range alignment of magnetic whirls, the so-called magneto-toroidal moments~\cite{Dubovik90, Gorbatsevich94, Ederer07, Spaldin08, Kopaev09, Gnewuch19}.
As shown in Fig.~\ref{Tisol}, these toroidal moments can be composed from an arrangement of elementary magnetic moments within the unit cell, or they may be exhibited monolithically by the elements themselves~\cite{Staub09, Scagnoli11, Spaldin13}, see Fig.~\ref{Tisol}.

\noindent
The uniform alignment of the magneto-toroidal moments $\mathbf{t}$ leads to a macroscopic toroidization $\mathbf{T}$ representing the order parameter, see Fig.~\ref{Tdom}.
The configuration of magnetic moments associated with the ferrotoroidic state breaks the space-inversion and time-reversal symmetries, with fundamental consequences for related electric and magnetic responses and couplings.
Specifically, magneto-toroidal materials allow for an electric-field-induced magnetization and a magnetic-field-induced polarization via the linear magnetoelectric effect~\cite{Fiebig05_R, Spaldin08, Thoele20, Watanabe18}.
Ferrotoroidic materials may thus be exploited for nanoelectronic memories or sensors that are based on intertwined magnetic and electric properties.
Furthermore, the linear magnetoelectric effect in the optical regime manifests as directional anisotropy and thus opens a pathway for photonic devices such as optical diodes~\cite{Rikken02, Arima08, Szaller13, Kezsmarki14, Toyoda15}.

\noindent
The scarcity of studies on imaging and manipulation of ferrotoroidic domains~\cite{Vanaken07, Zimmermann14} is largely due to the magnetically compensated nature of the ferrotoroidic state, which hampers experimental access to and thus a deeper understanding of the concept of magneto-toroidal order.
In addition, it is difficult to disentangle the fragile competition of exchange-interactions that is assumed to promote the toroidal order on the microscopic scale~\cite{Spaldin08, Lee14, Gnewuch19}.
To overcome these obstacles we take advantage of the definition of ferroic order as macroscopic phenomenon, irrespective of its explicit microscopic origin.
As shown in Refs.~\cite{Lehmann19, Lehmann20}, a transfer from the atomic to the sub-micrometer length scale, or mesoscale, provides a means to implement and probe ferrotoroidicity to a degree that conventional `atomic' materials cannot offer.
For these studies, the quantum-mechanical magnetic moments of a hypothetical magneto-toroidal crystal were replaced by classical macrospins in the form of magnetic single-domain sub-micrometer-sized permalloy bars, that can be lithographically patterned and arranged at will, thus allowing versatile tailoring of the symmetry and microscopic interactions of the resulting array.

\noindent
In this article, we build on studies of conjugate-field poling~\cite{Lehmann19} and manipulation of short- and long-range order~\cite{Lehmann20}.
These studies were performed on a single and very specific type of magneto-toroidal array.
We now present a variety of mesoscale magnetic systems, either comprised of single-domain bars or of equilateral triangles, as two fundamentally different types of building blocks for magneto-toroidal arrays.
While in the former case a ring-like arrangement of the magnetic-dipole-like building blocks~\cite{Harris10} exhibits a composite magneto-toroidal moment, the latter hosts a monolithic magneto-toroidal moment in each individual building block~\cite{Udalov12, Krutyanskiy13}.
Using magnetic-dipole calculations and micromagnetic simulations we quantify and tailor two variants of microscopic interactions that are required to promote ferrotoroidic ordering in these two types of systems --- an intra- and an inter-toroidal coupling.
Using magnetic force microscopy (MFM) we confirm the existence of as-grown magneto-toroidal domains, and we resolve the domain-wall states in our nanomagnetic structures.

\noindent
The article is organized as follows:
In Section~\ref{meth} we explain the process of fabricating and probing of our magneto-toroidal arrays, as well as our technique of micromagnetic simulations.
The composite and monolithic types of magnetic building blocks that provide the basis for the spontaneous formation of magneto-toroidal order are introduced in Section~\ref{model}.
We quantify the microscopic interactions between the building blocks and demonstrate the implementation of suitable couplings in arrays of composite and monolithic magneto-toroidal moments in Section \ref{int1} and \ref{int2}, respectively.
In Section~\ref{results} we present and discuss the experimental data on magneto-toroidal domains and domain-wall configurations.
We summarize our findings and put them into the larger context of magnetically compensated ferroic order in Section~\ref{conclusion}.

\section{Methods}
\label{meth}

\subsection{Sample fabrication}

\noindent
Arrays of sub-micrometre-sized building blocks made from ferromagnetic permalloy (Ni$_{81}$Fe$_{19}$) were fabricated using electron-beam lithography and deposition at room temperature.
To prepare for the patterning, a polymethyl methacrylate layer (2\% PMMA 950k) was spin-coated onto a 500-$\mu$m-thick (100)-oriented silicon substrate.
An electron-beam writer (Vistec EBPG 500Plus) operating with an acceleration voltage of 100\,kV at a dose of about 600\,$\mu$C/cm$^2$ was used to write the pattern into the PMMA resist.
After development, permalloy with thicknesses between 12 and 20\,nm was deposited via electron-beam evaporation at a growth rate of 0.3\,nm min$^{-1}$ at a base pressure of $10^{-6}$\,mbar.
The polycrystalline permalloy film with its negligible magneto-crystalline anisotropy ensures a distribution of the local magnetization within each building block that is determined primarily by its shape.
A thin capping layer of a few nanometer gold or aluminum was deposited on top of the permalloy to prevent deterioration due to permalloy oxidation.
Afterwards, remaining resist and unwanted material was removed via ultrasound-assisted lift-off in Technistrip~P1316.
The resulting arrays have lateral sizes of about $50\times50\,\mu$m$^2$.

\subsection{Micromagnetic imaging}

\noindent
To reveal the pristine, as-grown magneto-toroidal configuration in our arrays, we performed MFM (NT-MDT NTegra-Prima) using a two-pass mode with 40 to 50\,nm lift height.
We used tips with a low magnetic moment (Nanosensors PPP-LM-MFMR) to minimize the influence of the MFM tip onto the magnetic configuration of the scanned array.

\subsection{Micromagnetic simulations}

\noindent
The magnetic configuration and stray fields of the triangle-shaped building blocks, see Section~\ref{model}, were simulated using the program MuMax3~\cite{Vansteenkiste14}.
The geometric parameters of a single building block were discretized into a grid with cells of $2 \times 2 \times 4$\,nm$^3$ (length $\times$ width $\times$ height).
To simulate the magnetic configuration of the triangle, we use a saturation magnetization of $M_\text{sat}=860$\,kA\,m$^{-1}$, an exchange stiffness of $A_\text{ex}$=13\,pJ\,m$^{-1}$, and vanishing anisotropy ($K=0$).
The magnetic stray field is calculated for a single triangle-shaped building block surrounded by vacuum.

\section{Description of the model systems}
\label{model}

\begin{figure}[H]
	\centering
	\includegraphics[width = .34\textwidth]{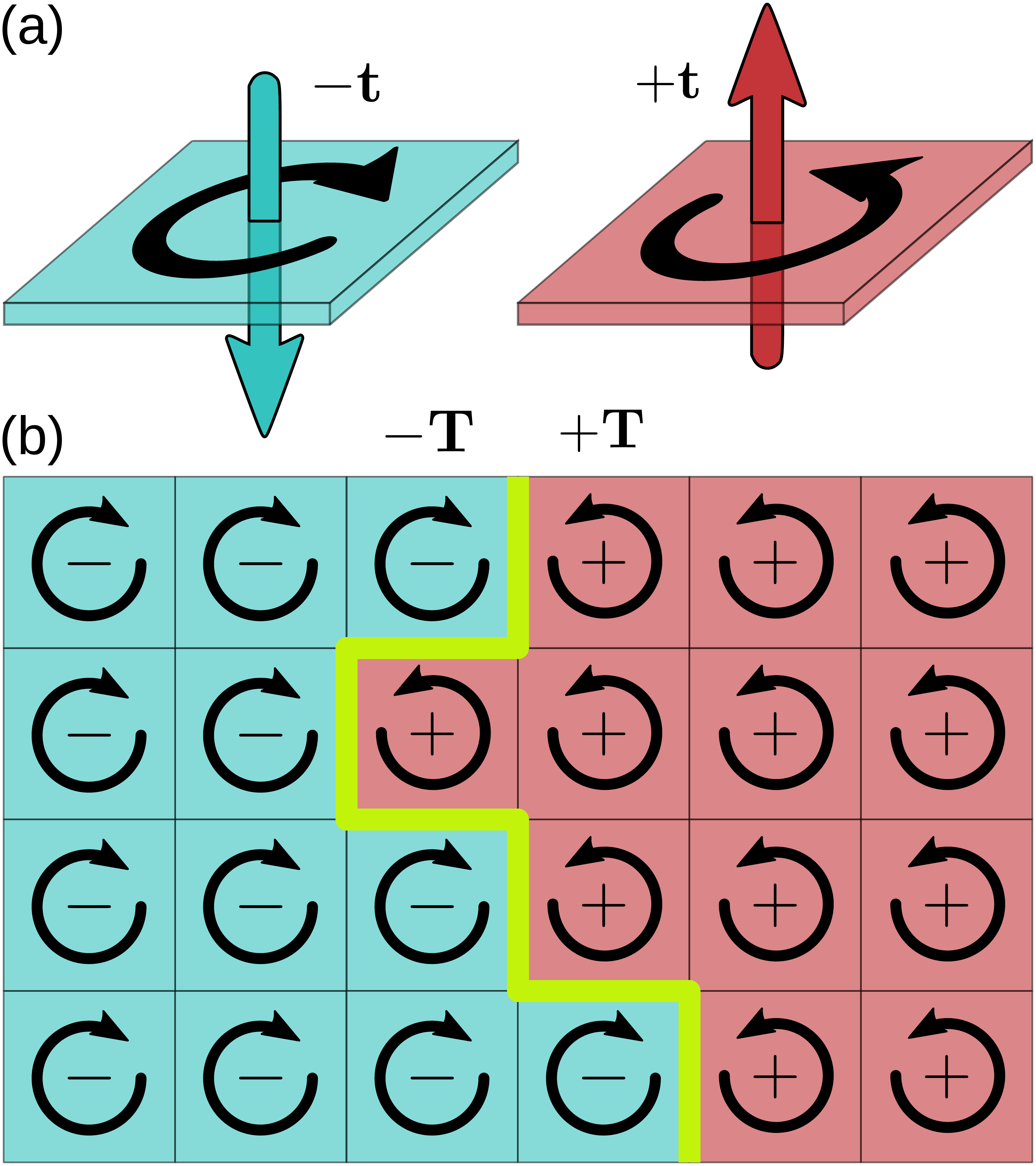}
	\caption{\label{Tdom}\textbf{Magneto-toroidal domain structure.}
	(a) Representation of the two magneto-toroidal-moment orientations (cyan: $-\mathbf{t}$, red: $+\mathbf{t}$) with magnetic moments in a clockwise or counter-clockwise head-to-tail orientation (black circular arrow).
	(b) Magneto-toroidal domain structure in which a domain wall (green line) separates states with a toroidization of $-\mathbf{T}$ (cyan) and $+\mathbf{T}$ (red).}
\end{figure}

\noindent
For implementing magneto-toroidal order in two-dimensional arrays of nanomagnets, two design criteria have to be considered.
First, the artificial unit cells of the arrays have to exhibit a magneto-toroidal moment as the basis of the macroscopic order.
This toroidal moment can either be formed as a composite of $n$ magnetic moments $\boldsymbol{\mu}_i$ located at positions $\mathbf{r}_i$ in the unit cell, see Fig.~\ref{Tisol}(a) ($\mathbf{t} \propto \sum_{i=1}^n \mathbf{r}_i \times \boldsymbol{\mu}_i$) or it originates from a continuous magnetic curl ($\mathbf{t} \propto V^{-1} \int_\text{uc} \mathbf{r} \times \boldsymbol{\mu}(\mathbf{r}) \, d^3r$ with $V$ as the volume of the entity and `uc' denominating the unit cell) forming a monolithic toroidal moment, see Fig.~\ref{Tisol}(b).
Second, since toroidal order refers to the spontaneous uniform alignment of these toroidal moments, a non-zero net toroidization $\mathbf{T} = N^{-1} \sum_{j=1}^N \mathbf{t}_j$, with $N$ as the number of unit cells contributing to the uniform alignment, has to emerge as the corresponding order parameter.
This leads to the formation of toroidal domain patterns, as sketched in Fig.~\ref{Tdom}.
Note that the transition from spins of the ions in conventional crystals to macrospins of the magnetic nanobars in our mesoscale arrays implies the replacement of the quantum-mechanical exchange interaction with the classical magnetic-dipole interaction between the building blocks~\cite{Luttinger46, Kraemer12, Alkadour17}.

\begin{figure}[H]
	\centering
	\includegraphics[width = .26\textwidth]{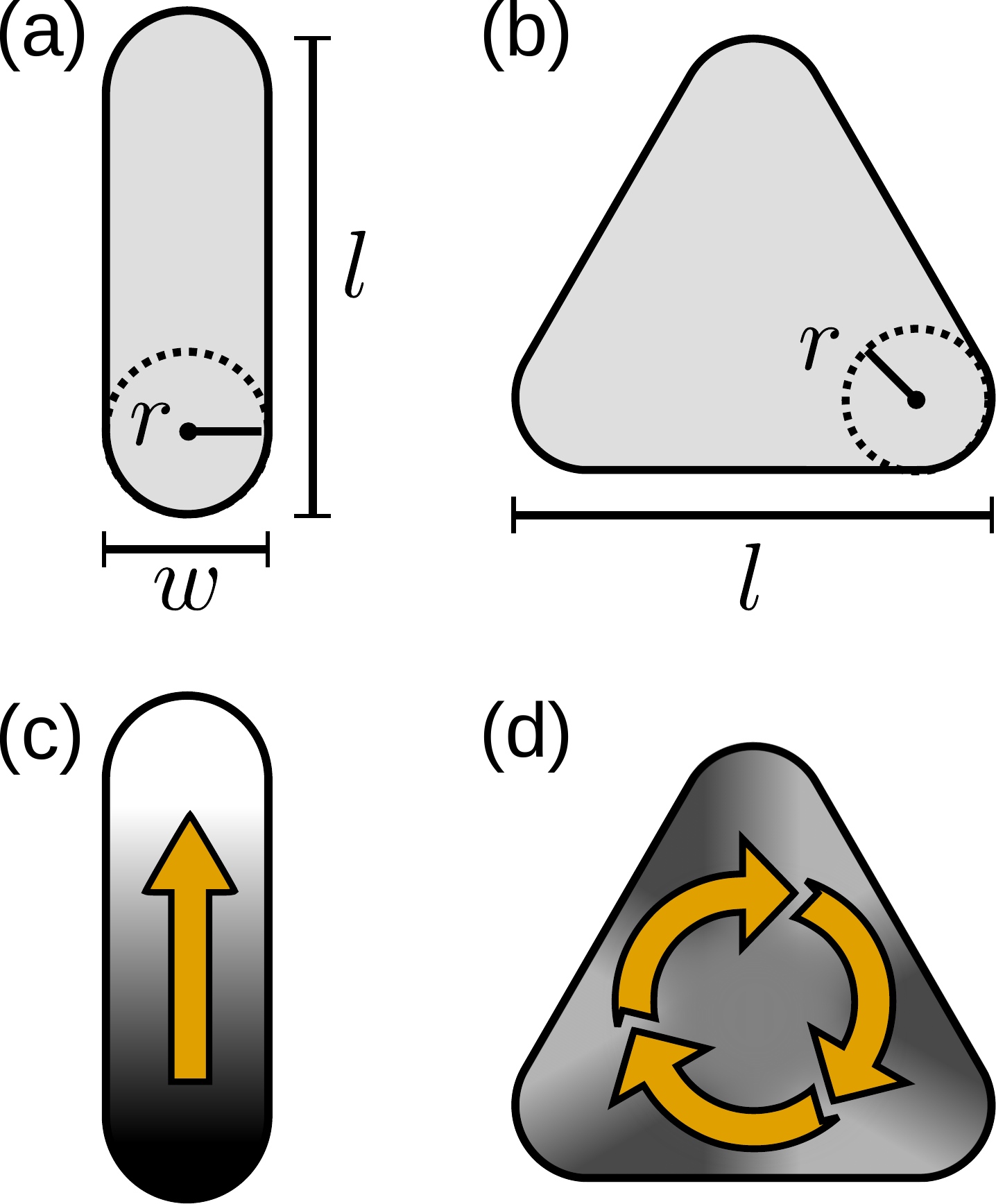}
	\caption{\label{iv}
		\textbf{Ferromagnetic constituents of composed and monolithic toroidal moments.}
		Our arrays comprise of (a) nanobars and (b) equilateral planar triangles arranged in different tilings, see Figs.~\ref{coup} and \ref{vortex}.
		Adjustable parameters are the length $l$, the width $w$, the curvature radius $r$, and the height $h$.
		(c) The nanobars exhibit a magnetic single-domain state with up- or down magnetization (orange arrow and gradient) as Ising-like degree of freedom.
		(d) The equilateral triangles exhibit a clockwise or counter-clockwise oriented magnetic vortex associated with a down- or up toroidal moment.}
\end{figure}

\noindent
In this work, we investigate mesoscale magnetic arrays based on two complementary types of building blocks promoting magneto-toroidal order.
These building blocks are made of ferromagnetic permalloy exhibiting a negligible magnetocrystalline anisotropy so that the shape of the nanoscale element determines its internal magnetic structure~\cite{Cowburn00}.
The first set of arrays consists of stadium-shaped bars, each with a size of $l=450$\,nm and $w=150$\,nm and a corner radius of $r=w/2=75$\,nm, see Fig.~\ref{iv}(a).
The bars carry an in-plane single-domain magnetization along their long axis, see Fig.~\ref{iv}(c).
Hence, these Ising-like macrospins represent a classical analogue of quantum-mechanical spins~\cite{Bedanta09}.
Such macrospins have been successfully used to address fundamental questions about magnetic correlations, frustration, thermal relaxation, phase transitions, and many other aspects~\cite{Skjaervo19}.

\noindent
The second set of arrays consists of equilateral triangles with $l=400$\,nm edge length and $r=,50$\,nm corner radius, see Fig.~\ref{iv}(b).
The appropriate choice of size, corner radius and thickness allows us to promote the formation of a magnetic vortex~\cite{Vogel12} as indicated in Fig.~\ref{iv}(d).
The triangular shape breaks the in-plane rotational symmetry and, in contrast to circular-shaped building blocks, supports non-zero magnetic stray fields~\cite{Yakata10, Udalov12, Krutyanskiy13}.
The stray field emanating from each triangle facilitates their magnetostatic coupling, which is required for the emergence of spontaneous long-range order.
In addition, the stray fields allow for the detection of the magnetic configuration by MFM.

\noindent
All the arrays presented in this work do not exhibit a net magnetization, so that a decomposition into an uncompensated (magnetized) and a compensated (toroidal) part of the spin arrangements as described in Ref.~\cite{Ederer07} is not necessary.
Furthermore, unlike in conventional ionic crystals, a phase transition promoting a macroscopic toroidization via a structural distortion \cite{Ederer07} is not possible.
Therefore, the crystal structure and the associated magnetic-moment configuration as such have to break space-inversion and time-reversal symmetries.

\noindent
Considering one of the key aspects of ferroic materials -- the reversibility of the order parameter in an external field -- non-centrosymmetric arrays of triangle-shaped magneto-toroidal elements, see Fig.~\ref{vortex}(c), offer interesting possibilities.
Such arrays facilitate the controlled reversal of the toroidization by the mere application of a homogeneous magnetic field due to the asymmetric nucleation energy of the vortex core~\cite{Thevenard10, Yakata10}.
This feature is a striking advantage in terms of applications based on the array's net toroidization.\\

\noindent
To identify arrangements of magnetic building blocks that promote a magneto-toroidal ordering, we limit our consideration to the magnetostatic interaction between direct neighbors.
This approximation is justified due to the rapid decrease of the magnetic stray field with increasing distance.
We distinguish between two types of couplings:
First, the distribution of magnetic stray fields within each basic unit has to favor a compensated, whirl-like configuration of magnetic moments, a condition we denominate as `intra-toroidal coupling'.
Second, the magnetic stray fields exhibited by these magneto-toroidal building blocks have to promote a parallel orientation of adjacent toroidal moments, a condition we denominate as `inter-toroidal coupling'.
For the arrays of composite and monolithic magneto-toroidal moments, these two types of coupling have fundamentally different origins with consequences for the emergent domain structure in the arrays, as we will explain in more detail.

\subsection{Microscopic interactions in arrays of composite magneto-toroidal moments}
\label{int1}

\begin{figure}[H]
	\centering
	\includegraphics[width = \textwidth]{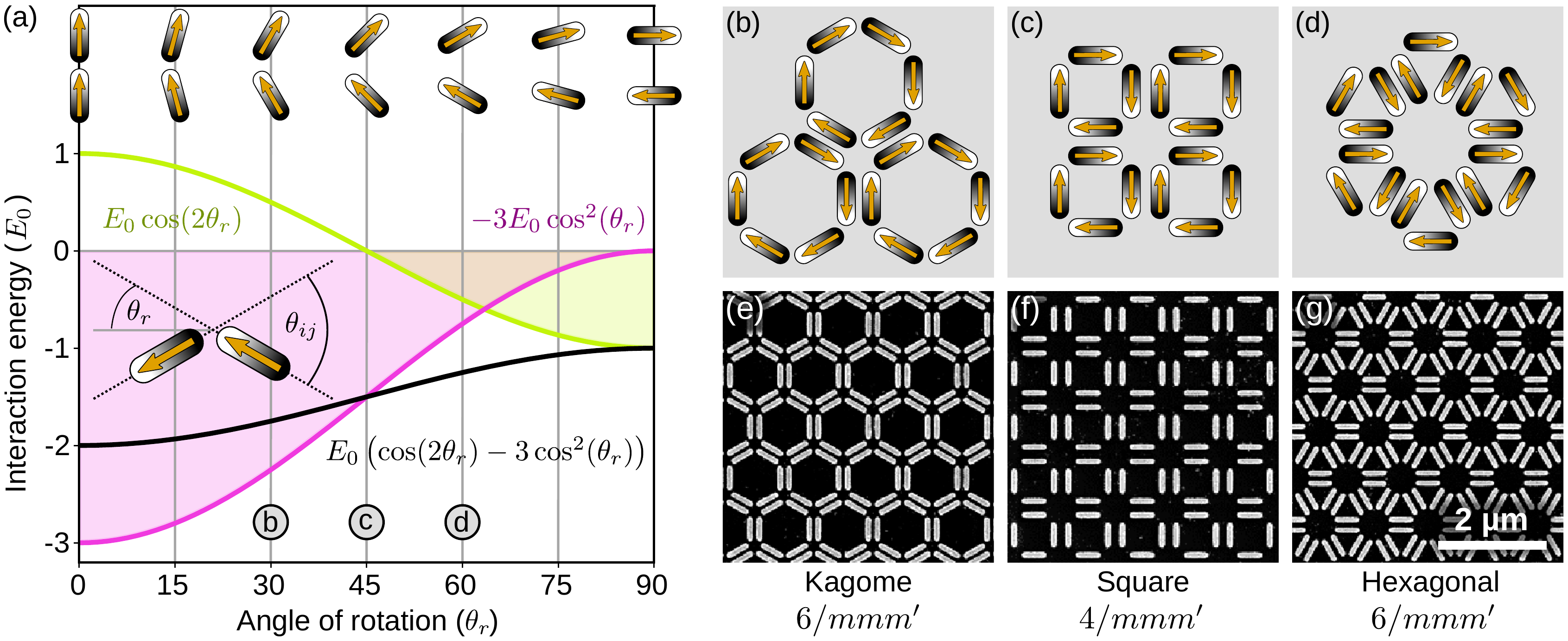}
	\caption{\label{coup}
	\textbf{Magnetic coupling in arrays of composite toroidal moments.}
	(a) Angle-dependent interaction energies, see Eq.~\ref{ed-angle}, for pairs of magnetic moments as sketched in the upper part of the panel.
	The purple and green curves represent the two types of couplings in Eq.~\ref{ed-angle} and quantify the contribution of the coupling terms to the total magnetic-dipole interaction energy (black curve).
	(b)~--~(d): Sketches of a kagome lattice at $\theta_r$=30° (b), a square lattice at $\theta_r$=45° (c), and a hexagonal lattice at $\theta_r$=60° (d) in a $-\mathbf{t}$ configuration with the specified magnetic point-group symmetry.
	(e)~--~(g): Corresponding scanning electron microscopy images showing sections of the three arrays.
	Arrays (b) and (d) comprise of permalloy bars of size $l=450$\,nm, $w=150$\,nm, and $h=12$\,nm on silicon, whereas the permalloy height of building blocks in array (c) is 20\,nm.
	The scale bar is the same for the arrays (b)--(d).}
\end{figure}

\noindent
For the arrays of composite magneto-toroidal moments, both the intra and inter-toroidal couplings originate from the magnetic-dipole-like stray fields generated by the magnetic single-domain bars.
The coupling energy $E_\text{D}$ of two interacting bars can be approximated by the magnetic dipole-dipole interaction between two point dipoles according to

\begin{equation}
E_\text{D} = \frac{\mu_0}{4\pi}  \left( \frac{\mathbf{m}_i \cdot \mathbf{m}_j}{|\mathbf{r}_{ij}|^3} - \frac{3  (\mathbf{m}_i \cdot \mathbf{r}_{ij}) (\mathbf{m}_j \cdot \mathbf{r}_{ij})}{|\mathbf{r}_{ij}|^5} \right) \; ,
\label{ed}
\end{equation}

\noindent
where $\mu_0$ is the vacuum permeability, $\mathbf{m}_{i,j}$ are vectors of the $i$-th and $j$-th magnetic moment, and $\mathbf{r}_{ij}$ is the vector connecting the two.
Equation~(\ref{ed}) can be rewritten by considering just the angle $\theta_{ij} = \arccos \left[ (\mathbf{m}_i \cdot \mathbf{m}_j)/(|\mathbf{m}_i| \, |\mathbf{m}_j|) \right]$ between the directions of the two neighboring magnetic moments $\mathbf{m}_i$.
We here limit our consideration to arrays in which neighboring magnets are placed as depicted in the upper panel of Fig.~\ref{coup}(a) and with a fixed distance $|\mathbf{r}_{ij}|$ between their centers.
We can now express Eq.~(\ref{ed}) as a function of the angle $\theta_\text{r} = \arccos \left[(\mathbf{m}_{i} \cdot \mathbf{r}_{ij})/(|\mathbf{m}_{i}| \, |\mathbf{r}_{ij}|) \right] = \theta_{ij} / 2$ between a magnetic moment and the distance vector $\mathbf{r}_{ij}$ to its nearest neighbour, as

\begin{equation}
E_\text{D} =  E_0 \left[\cos(2\theta_\text{r})-3 \cos^2(\theta_\text{r})\right] \; ,
\label{ed-angle}
\end{equation}

\noindent
with $E_0 = (\mu_0 |\mathbf{m}_{i,j}|^2)/(4 \pi |\mathbf{r}_{ij}|^3)$.

\noindent
To construct composite magneto-toroidal moments from macrospins, we place them in a circular arrangement forming the unit cell of the array.
Here we choose arrangements made of three ($2\theta_\text{r}=120^\circ$), four ($2\theta_\text{r}=90^\circ$), or six ($2\theta_\text{r}=60^\circ$) magnets forming kagome, square, and hexagonal lattices, respectively, see Figs.~\ref{coup}(b)--\ref{coup}(d).
As shown in Fig.~\ref{coup}(a), the intra-toroidal coupling that stabilizes a whirl-like magnetic configuration within the unit cell is dominated [Figs.~\ref{coup}(b), \ref{coup}(d)] or even solely determined [Fig.~\ref{coup}(c)] by the second term of Eq.~(\ref{ed-angle}) (purple line).
In the same manner, the inter-toroidal coupling that connects the unit cells is represented by the first term in Eq.~(\ref{ed-angle}) (green line).
In all our arrangements, the inter-toroidal coupling manifests itself as the antiparallel alignment of magnetic moments from neighboring unit cells at $2\theta_r = 180^\circ$.

\subsection{Microscopic interactions in arrays of monolithic magneto-toroidal moments}
\label{int2}

\noindent
In contrast to the magnetic-dipole interaction determining the order in arrays of composite magneto-toroidal moments, the types of microscopic interactions promoting long-range order in arrays of monolithic magneto-toroidal moments are of a fundamentally different nature.
In particular, the intra- and inter-toroidal couplings have separate sources.
The intra-toroidal coupling results from competing contributions to the free energy within the individual ferromagnetic building block, which are primarily the magnetostatic interaction favouring flux-closed configurations of magnetic moments with minimized stray fields and the magnetic exchange interaction striving for a parallel and uniform spin alignment with a saturated magnetization.
While the design parameters of the nanobars shown in Fig.~\ref{iv}(a) can be tailored to obtain a uniformly magnetized macrospin-like configuration, see Fig.~\ref{iv}(c), the triangles in Fig.~\ref{iv}(b) can be engineered to stabilize a magnetic vortex configuration as shown in Fig.~\ref{iv}(d).

\begin{figure}[H]
	\centering
	\includegraphics[width = \textwidth]{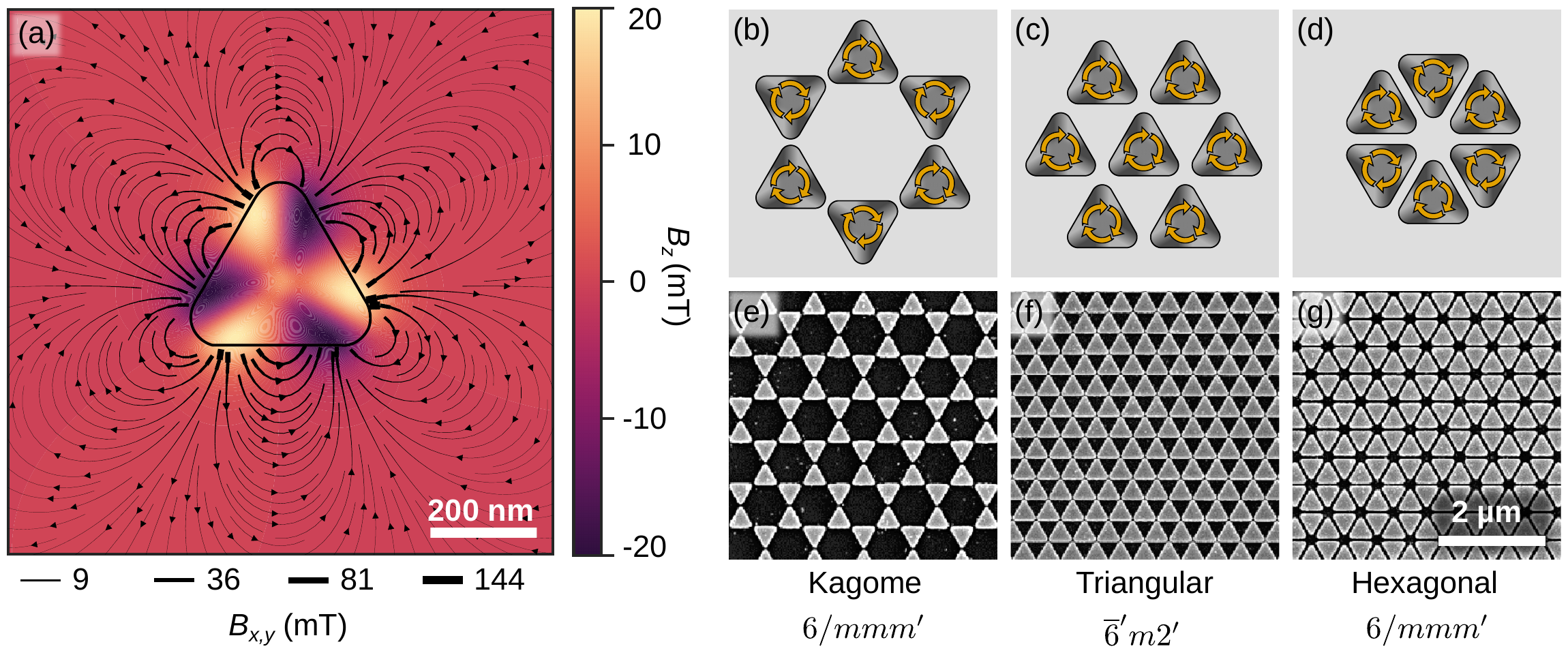}
	\caption{\label{vortex}
	\textbf{Magnetic coupling in arrays of monolithic toroidal moments.}
	(a) Micromagnetic simulation of the magnetic field surrounding a permalloy triangle (outlined) in its magnetic ground state with size $l=400$\,nm, $r=50$\,nm, and $h=20$\,nm.
	The color scale denotes the strength of the magnetic out-of-plane stray field 60\,nm above the substrate.
	The streamline plot indicates the in-plane magnetic-field strength $B_{x,y}$ at the triangles' plane.
	(b)~--~(d): Sketches of a kagome lattice (b), a triangular lattice (c), and a hexagonal lattice (d) in a $-\textbf{t}$ configuration with the specified magnetic point-group symmetry.
	(e)~--~(g): Corresponding scanning electron microscopy images showing sections of the three arrays.
	Arrays (e) and (g) comprise of 12-nm-thick permalloy triangles deposited on silicon, whereas array (f) is made from 20-nm-thick permalloy triangles.
	The scale bar is the same for the arrays (e)--(g).}
\end{figure}

\noindent
Instead of a dipole-like magnetic field surrounding the uniformly magnetized nanobars, the triangles exhibit an inherently weaker hexapole-like magnetic field as shown in Fig.~\ref{vortex}(a) as the basis of the inter-toroidal coupling.
The position of the six magnetic poles of alternating sign surrounding each triangle determine suitable arrangements that provide an inter-toroidal coupling.
We designed a variety of corner- and edge-coupled networks of magnetic triangles, forming kagome, triangular, and hexagonal lattices, respectively, see Figs.~\ref{vortex}(b)--\ref{vortex}(d).
The proximity of oppositely charged magnetic poles in our structures introduces a coupling that favors locally a parallel alignment of neighboring toroidal moments and, hence, a global toroidization.

\section{Experimental Results and Discussion}
\label{results}

\subsection{Long-range order in arrays of composite magneto-toroidal elements}

\begin{figure}[H]
	\centering
	\includegraphics[width = \textwidth]{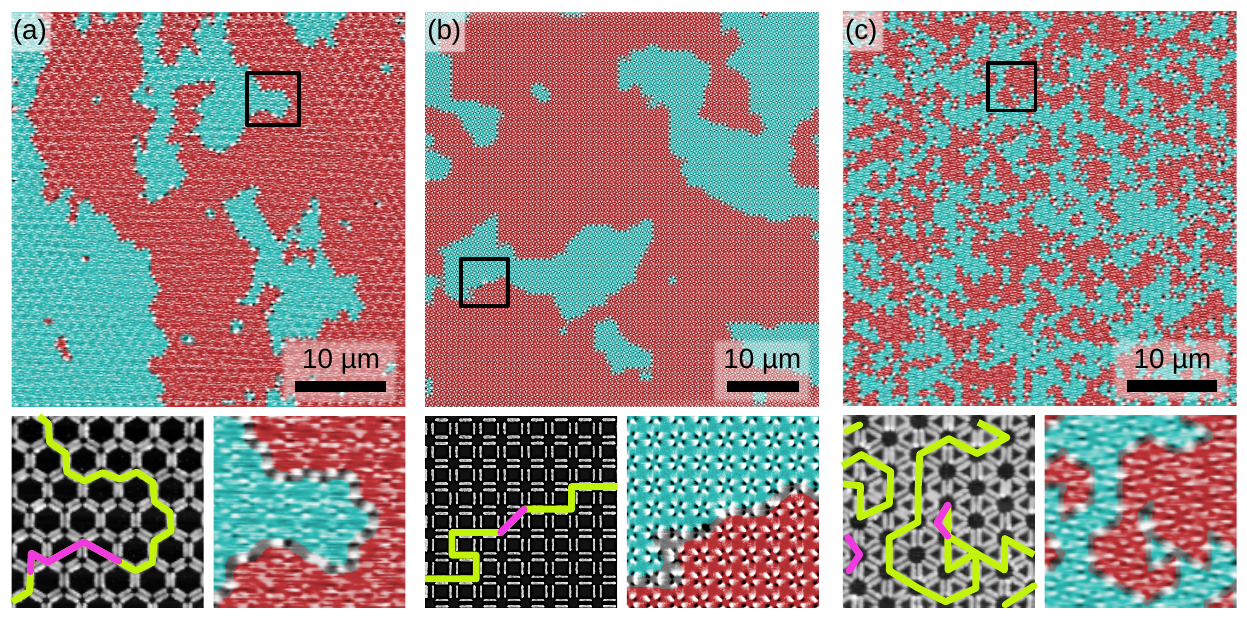}
	\caption{\label{Isingdom} \textbf{Long-range order in arrays of composite magneto-toroidal moments.}
	MFM scans with color-coded domain states (cyan: $-\mathbf{T}$, red: $+\mathbf{T}$) of the three arrays with composite magneto-toroidal moments as introduced in Fig.~\ref{coup}.
	Lower panels: Magnified topography (left) and corresponding magneto-toroidal order (right) of areas around a domain wall.
	Two types of domain-wall configurations are indicated that run either in between (inter-toroidal walls, green lines) or across (intra-toroidal walls, purple lines) the magneto-toroidal unit cells.}
\end{figure}

\noindent
Figure~\ref{Isingdom} shows the magneto-toroidal-domain configurations measured by MFM on the three macrospin-based arrays shown in Figs.~\ref{coup}(b)--\ref{coup}(d).
The MFM scans performed on as-grown arrays reveal the local toroidization as well as the structure of the walls separating areas with toroidization $-\mathbf{T}$ (cyan) and $+\mathbf{T}$ (red).
For all arrays we find spontaneous magneto-toroidal order with domains that extend laterally over a few to several tens of unit cells.
The formation of finite-sized domains can be regarded as a freezing-out of a non-equilibrium configuration during the growth of the arrays.
With the increasing permalloy-film thickness during growth, the coupling strength between neighboring elements increases such that thermal fluctuations get continuously suppressed.
As a consequence, the ongoing deposition emulates a continuous decrease of the systems' temperature, which eventually quenches the system from a superparamagnetic state through the symmetry-breaking phase transition into a non-equilibrium multi-domain configuration.
The average size of the domains as well as the microstructure and density of the domain walls is governed by the the domain-wall energy and the number of energetically degenerate domain-wall states.
Note that the three arrays were grown at different times so that a qualitative comparison of parameters such as the observed domain sizes would be impeded by systematic variations and is therefore of limited significance.

\subsection{Long-range order in arrays of monolithic magneto-toroidal elements}

\begin{figure}[H]
	\centering
	\includegraphics[width = \textwidth]{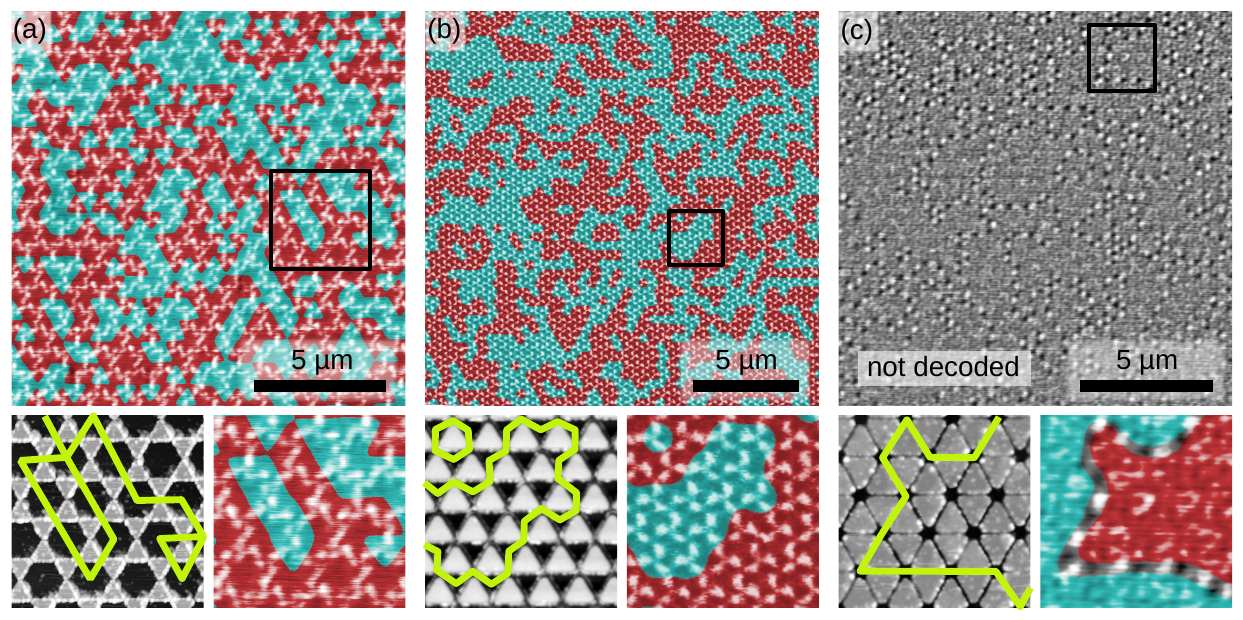}
	\caption{\label{Vortexdom} \textbf{Long-range order in arrays of monolithic magneto-toroidal moments.}
	MFM scans with color-coded domain states (cyan: $-\mathbf{T}$, red: $+\mathbf{T}$) of the three arrays with monolithic magneto-toroidal moments as introduced in Fig.~\ref{coup}.
	Lower panels: Magnified topography (left) and corresponding magneto-toroidal order (right) of areas around a domain wall.
	Due to the monolithic nature of the building blocks, only inter-toroidal walls emerge (green lines).}
\end{figure}

\noindent
The MFM scans performed on the three triangle-based arrays of Figs.~\ref{vortex}(b)--\ref{vortex}(d) are shown in Fig.~\ref{Vortexdom}.
The measurements reveal spontaneous magneto-toroidal order and domains that extend laterally across a few unit cells only.
For the kagome and the triangular arrays shown in Figs.~\ref{Vortexdom}(a) and \ref{Vortexdom}(b) we found that all triangles form a magnetic vortex state as sketched in Fig.~\ref{iv}(d).
Obviously, the intra-toroidal coupling dominates in the arrays of monolithic toroidal moments, stabilizing the vortex state in the triangles against the formation of energetically unfavorable uniformly magnetized configurations without a magneto-toroidal moment.
The different microscopic sources of the competing couplings, namely the interplay of magnetic exchange and magnetostatic interaction for the intra-toroidal coupling, as well as the magnetostatic interaction for the inter-toroidal coupling, apparently constitute a native dominance of the latter, which explains our observations.

\noindent
The dense packing of triangles in the hexagonal array in Fig.~\ref{Vortexdom}(c) reduces the out-of-plane magnetic stray fields and lowers the MFM contrast such that an unambiguous assignment of the toroidal domain structure works in selected areas only, as e.g. shown in the inset in Fig.~\ref{Vortexdom}(c).
In contrast, at the domain walls (see the highlighted area in Fig.~\ref{Vortexdom}(c)), adjacent triangles exhibiting an opposite toroidal moment yield enhanced out-of-plane magnetic stray fields that are well detectable by MFM.

\subsection{Domain walls and their substructure in magneto-toroidal arrays}
\label{walls}

\noindent
Domain walls can be regarded as local correlated excitations in ordered systems that originate from the reorientation of the order parameter when moving from one domain to another.
The study of domain walls is of fundamental interest as their presence and manipulability determines technological key parameters of ferroic materials, such as their `hardness' and transport properties.
Here, the transfer from atomic to mesoscopic magneto-toroidal systems enables unparalleled insights into the structure of the domain walls.
The walls in our magneto-toroidal arrays are highlighted in the lower panels of Figs.~\ref{Isingdom} and \ref{Vortexdom}.
We observe two types of walls, which either run in between or across the magneto-toroidal building blocks, as indicated in the lower panel of Fig.~\ref{Isingdom}, with green (inter-toroidal walls) and purple (intra-toroidal walls) lines, respectively.
The preferred type of domain-wall state is the one that requires the least amount of energy for its formation, which is determined by the relative strength of the microscopic couplings in the arrays.
The observed preference of inter-toroidal domain walls (green lines) points to a dominating intra-toroidal coupling for both types of arrays.
However, while in the arrays of composite magneto-toroidal moments both couplings stem from the magnetic-dipole interaction and are of comparable magnitude, the interactions in the arrays of monolithic magneto-toroidal moments result from different mechanisms with intrinsically different magnitudes, as described above.
Previous work from Adeyeye \textit{et.\,al.}~\cite{Adeyeye07} indicates that by considering the local Zeeman-energy contribution from the magnetic stray fields of vortex elements in densely packed mesoscale arrays, even the monolithic magnetic vortex state may destabilize in favor of a uniformly magnetized configuration at some point.
Hypothetically, intra-toroidal walls in arrays of monolithic magneto-toroidal elements may, hence, emerge as uniformly magnetized triangles at the domain wall.
Within the structural parameters chosen for our arrays, however, we did not observe this kind of state.\\

\noindent
If both domain-wall types are present in a sample, lower-dimensional domains within the domain walls become possible.
We found that the three arrays with composite magneto-toroidal moments display such a substructure of the domain walls, whereas no such substructure was found in the three arrays with monolithic magneto-toroidal moments as we solely observed inter-toroidal walls.

\noindent
Furthermore, it has been shown for the composite-type magneto-toroidal square array that the type of domain wall determines the net magnetization direction of the magnetic moments forming the wall~\cite{Lehmann20}.
As a consequence, the meeting points of two domain-wall types constitute local sinks and sources of magnetic flux and can be described as emergent magnetic charges of either sign~\cite{Hubert09} that we found in all our arrays of composed magneto-toroidal moments.

\section{Conclusion}
\label{conclusion}

\noindent
In conclusion, we studied the spontaneous uniform alignment of magnetic whirls, so-called magneto-toroidal moments, as a type of ferroic order.
Our experimental systems are arrays of nanoscale building blocks from a soft-magnetic alloy that we fabricated by electron-beam lithography and deposition.
With this transfer of the crystal structure and its interactions from the atomic scale to the mesoscale, we achieved an unparalleled local experimental access of the magneto-toroidal state.
We distinguish between arrays of composite and monolithic magneto-toroidal moments.
While the former type of array exhibits magneto-toroidal moments that consist of a ring-like arrangement of magnetic single-domain nanobars representing the classical analog to spins, the latter type consists of ferromagnetic triangles that host an intrinsic, monolithic magneto-toroidal moment.
These model systems are chosen to emulate spin arrangements in the unit cell of prototypical ferrotoroidics, and orbital currents that give rise to toroidal moments localized at particular ions in a crystal lattice, respectively.
Using macrospin calculations and micromagnetic simulations, we quantify the inter- and intra-toroidal couplings that promote the emergence of magneto-toroidal order in both types of nanomagnetic arrays.

\noindent
Using MFM we confirmed the emergence of spontaneous long-range order in our arrays with magneto-toroidal domains that span over a few to several tens of unit cells.
Our study reveals the presence of two types of domain walls in the arrays of composite magneto-toroidal moments.
The walls either run in between or across the toroidal building blocks, which is associated with a dominance of the intra- or inter-toroidal coupling, respectively.
In the arrays with composite magneto-toroidal moments both the intra- and inter-toroidal coupling are determined by the magnetic-dipole interaction.
Therefore, both couplings are of equal magnitude and both types of domain walls are observed.
In contrast, in the arrays with monolithic magneto-toroidal moments the intra-toroidal coupling, which is given by the interplay of the magnetic-dipole and the magnetic-exchange interaction, inherently dominates so that only inter-toroidal walls are observed.

\noindent
In a more general framework, we shed light on ferrotoroidicity as a new and elusive type of net-magnetization-free long-range magnetic order that spontaneously breaks space-inversion and time-reversal symmetries.
As a consequence, the ferrotoroidic state exhibits potential for the exploitation of its intrinsic linear magnetoelectric effect and unique nonreciprocal optical responses associated to it.
Both phenomena are of fundamental interest and could be useful in the development of new functional materials with possible future applicability in memory arrays, sensors, and photonic devices.
With the demonstration of spatially resolved magneto-toroidal order in mesoscale magnetic arrays our work displays the fundamental benefits of utilizing such classical systems for studying subtle and complex ordering phenomena.

\begin{acknowledgments}
	
\noindent
The authors thank Th.\ Lottermoser, A.\ Bortis, P.~M.\ Derlet and C.\ Donnelly for valuable discussions.
M.~F.\ acknowledges funding by the Swiss National Science Foundation (projects no.\ 200021-175926 and 200021-178825) and the European Research Council (advanced grant no.\ 694955 INSEETO).
J.~L.\ and M.~F.\ acknowledge funding by the ETH Research grant no.\ ETH-28~14-1.
N.~L.\ and L.~J.~H.\ acknowledge funding by the Swiss National Science Foundation (project no.\ 200021-155917).
N.~L. acknowledges funding by the European Union's Horizon 2020 research and innovation program (Marie Sklodowska-Curie grant no.\ 844304 LICONAMCO).
\end{acknowledgments}

\bibliography{lib}

\end{document}